\documentclass[11pt]{article}
\usepackage{moriond,epsfig}
\usepackage{capt-of}

\newcommand{\pt}{p_{T}}
\newcommand{\Et}{E_{T}}
\newcommand{\MET}{\mbox{$E\kern-0.50em\raise0.10ex\hbox{/}_{T}$}}
\newcommand {\GeVc} {~\rm{GeV}\!/c}

\newcommand {\GeV} {~\rm{GeV}\!/c^2}

\newcommand {\pb} {~\rm{pb}^{-1}}
\newcommand {\fb} {~\rm{fb}^{-1}}

\newcommand{\nunubar}{\nu \overline{\nu}}
\newcommand{\llnn}{l^+l^-\nunubar}
\newcommand{\ppbar}{p \overline{p}}

\newcommand{\ttbar}{t \overline{t}}
\newcommand{\MetDeltaPhi}{\Delta\phi(\MET, \rm{lepton, jet})}
\newcommand{\METsig}{\mbox{$E\kern-0.50em\raise0.10ex\hbox{/}_{T,\mathrm{sig}}$}}
\newcommand{\METrel}{\mbox{$\min E\kern-0.50em\raise0.10ex\hbox{/}_{T,\mathrm{rel}}$}}
\newcommand{\SumEt}{\sum{E_{T}}}

\begin{document}
\vspace*{4cm}
\title{Searching for $H\rightarrow WW^*$ and Other Diboson Final States at CDF}

\author{ Shih-Chieh Hsu \\
(for the CDF Collaboration)}

\address{Department of Physics, University of California, San Diego, \\
La Jolla, California 92093}

\maketitle\abstracts{
We report searches for standard model (SM) Higgs production decaying to $WW^{(*)}$
and continuum $ZZ$ production in the two charged lepton and two neutrino final states.
The data were collected with the CDF II detector at the Fermilab Tevatron and 
correspond to an integrated luminosity of $1.1\fb$. In order to separate the processes contributing
to the final state, event probabilities calculated using the leading order differential
cross-sections were used to construct a likelihood ratio discriminant.
The observed (median expected) 95\% C.L. upper limit for $\sigma(H\rightarrow WW^{(*)})$ 
with 160$\GeV$ mass 
hypothesis is 1.3(1.8) pb which corresponds to 3.4(4.8) times the SM prediction at 
next-to-next-to-leading logarithmic level (NNLL) calculation.\cite{Catani:2003zt} 
The significance of the observed $ZZ$ signal is 1.9 $\sigma$ and the 95\% C.L.
upper limit is 3.4$\pb$ which is consistent with the next-to-leading order (NLO) calculation
of $1.4\pm0.1\pb$.
}

%%%%%%%%%%%%%%%%%%%%%%%%%%%%%%%%%%%%%%%%%%%%%%%%%%%%%%%
%                     Introduction                    %
%%%%%%%%%%%%%%%%%%%%%%%%%%%%%%%%%%%%%%%%%%%%%%%%%%%%%%%
\section{Introduction}

The Higgs boson is introduced into the standard model (SM) to explain 
electroweak symmetry breaking and the origins of particle mass.
Precision electroweak measurements and direct
searches have constrained the Higgs mass to lie between 114 and 182 $\GeV$ at the 95\% C.L.\cite{LEPEWWG}
We search for the Higgs boson through the gluon fusion production and decay channel,
$gg\to H \to WW^{*}$, which is the dominant channel for a Higgs with $m_H>135\GeV$. 
The maximum Higgs cross-section times branching fraction
for the $\ppbar \rightarrow H \rightarrow WW^*$ process is $0.388 \pb$ at NNLL and occurs at
the mass $m_H = 160\GeV$.
This is a small signal compared to continuum $WW$ production
which has a cross-section of $12.4\pb$ at NLO\cite{MCFM}.
A good understanding of the SM diboson production is essential for this search.
To get a good signal to background ratio sample, we search for fully leptonic decay 
of $WW^{*}\to l^+l^-\nu\bar{\nu}$,
where $l^{\pm}$ = e, $\mu$ or $\tau$ and $\tau$ decays to e or $\mu$. 
Pair production of Z bosons also decays to the same final state and has not
yet been seen at a hadron collider	 
The analysis strategy is to 
maximize the signal acceptance by loosing selection cuts and use the likelihood ratio discriminator(LR) 
calculated by Matrix Element methods to set the limits for 10 different Higgs mass hypotheses and
to search for ZZ production.

%%%%%%%%%%%%%%%%%%%%%%%%%%%%%%%%%%%%%%%%%%%%%%%%%%%%%%%
%                  Event Selection                    %
%%%%%%%%%%%%%%%%%%%%%%%%%%%%%%%%%%%%%%%%%%%%%%%%%%%%%%%
\section{\label{sec:Selection} Selection}

The $\llnn$ candidates are selected from two opposite-sign leptons from the same vertex
and high missing transverse energy $\MET$. At least one 
lepton is required to satisfy the trigger and have  $\pt > 20 \GeVc$. The other lepton
has looser requirement $\pt > 10 \GeVc$ to increase the kinematic acceptance. 
This sample receives contributions from continuum $WW, WZ, ZZ, t\bar{t},$
Drell-Yan, and $W\gamma$  and $W$+jets where the $\gamma$ or jet is misidentified as a lepton.
To suppress the $W$+jets background, we require leptons to be both track and calorimeter isolated
such that the sum
of the $\Et$($\pt$) for the calorimeter towers (tracks) in a cone of $\Delta R =
\sqrt{(\Delta\eta)^2 + (\Delta\phi)^2} < 0.4$ around the lepton is
less than 10\% of the $\Et$ for electrons or $\pt$ for muons and track
lepton candidates.
To suppress the Drell-Yan background, we require $\METrel>$ 25 GeV,
where $\METrel$ is defined to be:
  \begin{equation}
     \METrel \equiv \left\{ 
     \begin{array}{ll} 
     \MET                       & \mbox{ if } \MetDeltaPhi > \frac{\pi}{2} \\
     \MET\sin({\MetDeltaPhi})   & \mbox{ if } \MetDeltaPhi < \frac{\pi}{2} \\
     \end{array} \right.
  \end{equation} 
This definition will reject events whose observed $\MET$ is consistent with
the mis-measurement  of  a single jet or lepton in the event	 
We further require the candidates to have
less than 2 jets with $\pt>$ 15 GeV and $|\eta| < 2.5$, in order to suppress $\ttbar$ backgrounds,
$M_{\ell^{+}\ell^{-}}>$  25 GeV in order to suppress heavy flavor contributions, and
exactly 2 leptons to suppress $WZ$ contributions with a third lepton.

For the $ZZ$ analysis, the $e\mu$ channel is not used and one additional cut, 
$\METsig \equiv \MET / \sqrt{\SumEt}>$ 2.5 GeV$^{\frac{1}{2}}$, where ${\SumEt}$ is the scalar sum 
of calorimeter transverse energy, is applied to further suppress the
effect of mis-measurement of unclustered energy.

%%%%%%%%%%%%%%%%%%%%%%%%%%%%%%%%%%%%%%%%%%%%%%%%%%%%%%%
%                  Methods                            %
%%%%%%%%%%%%%%%%%%%%%%%%%%%%%%%%%%%%%%%%%%%%%%%%%%%%%%%

\section{\label{sec:MatrixElement} Event Probability Calculation}

In order to use all the kinematic information  available in the event to distinguish the modes
        contributing to the selected sample, we use an 
event-by-event calculation of the probability density function $P_{m}(x_{obs})$ for a mode $m$ 
which is either Higgs, $WW$, $ZZ$, $W\gamma$ or $W$+parton:
\begin{equation}
 P_{m}(x_{obs}) = {\frac{1} { < \sigma_{m} >} } \int {d \sigma^{th}_{m} (y) \over dy} \epsilon(y) G(x_{obs},y) dy 
\label{eqn:MEcalc}
\end{equation}
where 
$x_{obs}           $ are the observed lepton four-vectors and $\vec{\MET}$,
$y	           $ are the true lepton four-vectors (include neutrinos),
$  \sigma^{th}_{m} $ is the MCFM\cite{Campbell:1999ah} leading-order theoretical calculation of the cross-section for mode $m$,
$ \epsilon(y)      $ is total event efficiency $\times$ acceptance, 
$ G(x_{obs},y)     $ is an analytic model of resolution effects, and
$ \frac{1} { < \sigma_{m} > }$ is the normalization.
The function $\epsilon(y)$ describes the probabilities of a parton level object 
(e, $\mu$, $\gamma$ or parton) to be reconstructed as an observed lepton and
is extracted from a combination of Monte Carlo and data.
The event probability density functions are used to construct a dimensional discriminator:

\begin{equation} 
LR(x_{obs})\equiv\frac{P_{H}(x_{obs})}{P_{H}(x_{obs})+\Sigma_i k_i P_{i}(x_{obs})},
\end{equation}
where $H$ is Higgs, $k_i$ is the expected fraction for each background and $\Sigma_i k_i=1$.
For SM $ZZ$ search, we just use $ZZ$ and $WW$ to construct the discriminator $P_{ZZ}/(P_{ZZ}+P_{WW}).$

%%%%%%%%%%%%%%%%%%%%%%%%%%%%%%%%%%%%%%%%%%%%%%%%%%%%%%%
%                     Systematics                     %
%%%%%%%%%%%%%%%%%%%%%%%%%%%%%%%%%%%%%%%%%%%%%%%%%%%%%%%

\section{Systematics}

The trigger efficiency uncertainty (0.3\%$-$0.6\%) is measured from data.
The $\MET$ resolution modeling uncertainty (1\%$-$20\%) and lepton identification uncertainty (1.4\%$-$1.8\%) 
are determined
from comparisons of the data and the Monte Carlo simulation in a
sample of dilepton events.
For the $W\gamma$ background contribution, there is an additional uncertainty
of 20\% from the detector material description and conversion veto efficiency. 
The higher order effects in $WW$ (4.5\%) is assigned to be a half of the difference between 
the Pythia and {\sc MC@NLO}\cite{Frixione:2002ik} acceptance. 
The theoretical cross-section uncertainties (10\%$-$15\%) are assigned from NLO calculation.
The Parton Density Function uncertainties (1.9\%$-$2.7\%) are the quadrature sum of variations
between CTEQ5L and CTEQ6M. 
The systematic uncertainty of the $W$+jets background estimate is determined
to be 26.8\% from the dependence on the sample selection in the measurement of the rate at
which a jet is  misidentified as a lepton.
An additional 6\% uncertainty originating from the luminosity measurement 
is assigned to both signal and background except W+jets.

%\begin{table}[ht]
%\tiny
%\caption{The systematics for $l^+l^-\MET$ analysis. The numbers in the parenthesis are for the ZZ search.}
%\begin{center}
%\input{syst_Base}
%\label{tab:syst_llnn}
%\end{center}
%\end{table}

%%%%%%%%%%%%%%%%%%%%%%%%%%%%%%%%%%%%%%%%%%%%%%%%%%%%%%%
%                     HWW                             %
%%%%%%%%%%%%%%%%%%%%%%%%%%%%%%%%%%%%%%%%%%%%%%%%%%%%%%%
\section{$H\rightarrow WW^*$ Results}

\begin{figure}[ht]

\begin{minipage}[b]{0.7\columnwidth}%
    \centering
    \includegraphics[width=0.475\textwidth]{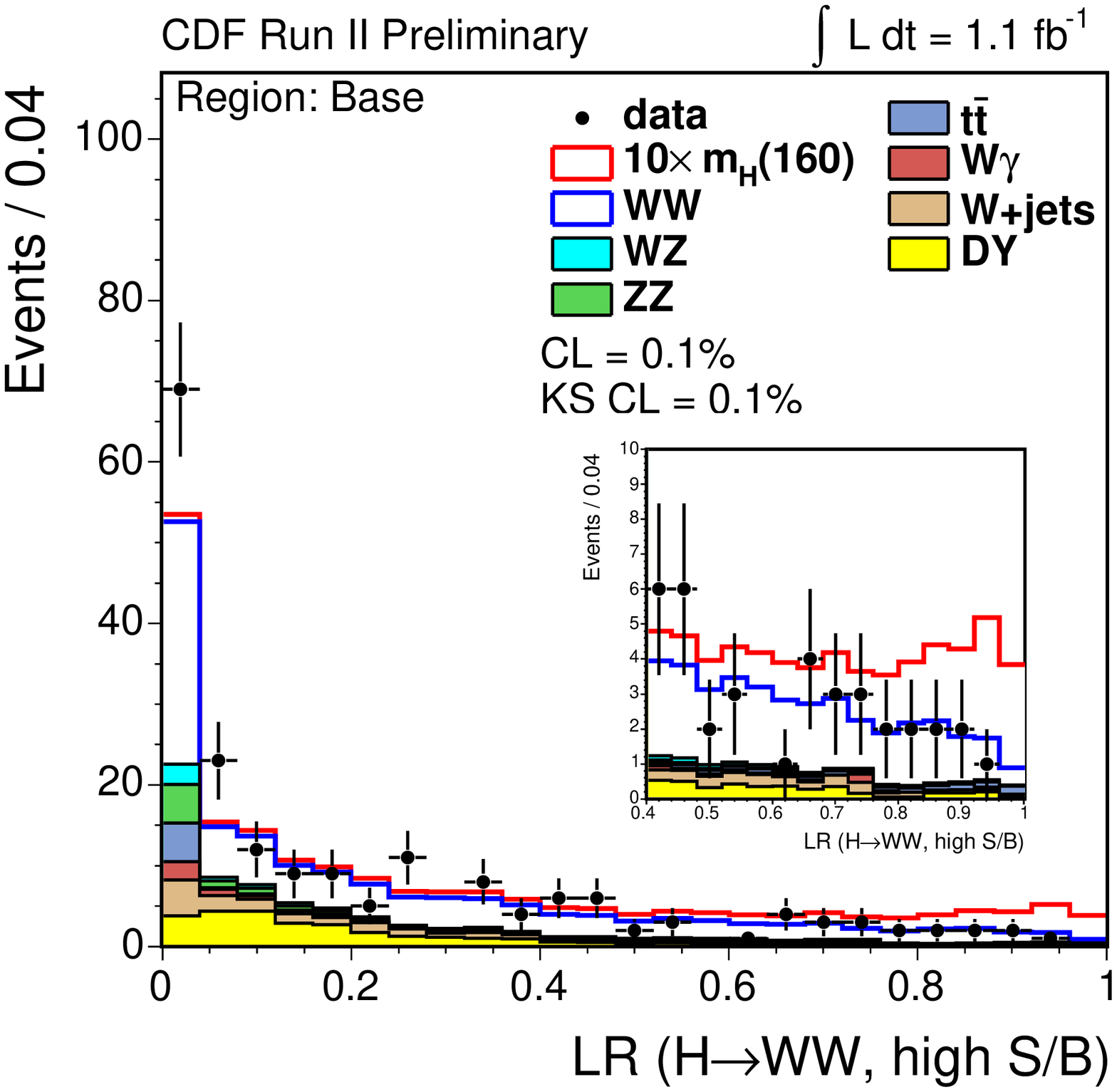}
    \unitlength=0.5\linewidth
    \put(-0.72,0.74){(a)}
    \includegraphics[width=0.475\textwidth]{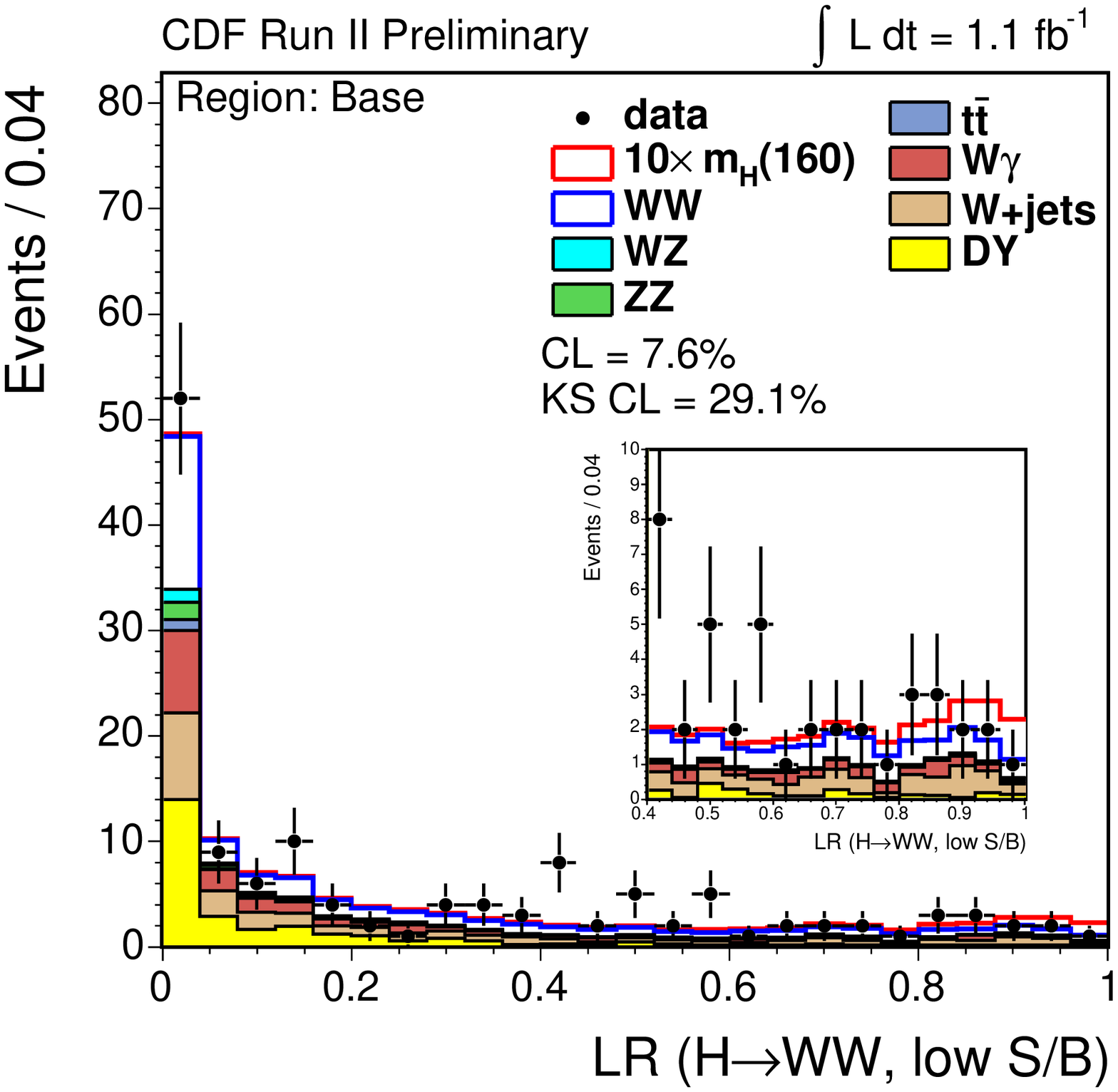}
    \put(-0.72,0.74){(b)}
    \caption{The LR distributions of Higgs mass 160 GeV$/c^2$ for (a) High S$/$B channel and (b) Low S$/$B channel. }\label{fig:HWW_LR}
\end{minipage}%
\hfill%
\begin{minipage}[b]{0.25\columnwidth}%
\begin{tabular}{|c|c|}
\hline
 & \small Expected \\ \hline
\small $ WW$        &  \small $132.9 $ \\
\small $ WZ$        &  \small $9.5   $ \\
\small $ ZZ$        &  \small $11.7  $ \\
\small $ t\bar{t}$  &  \small $9.6   $ \\
\small $ DY$        &  \small $55.4  $ \\ 
\small $ W\gamma$   &  \small $24.7  $ \\
\small $ W$+jets    &  \small $42.4  $ \\ \hline
\small Total        &  \small $286.1 \pm  23.3$ \\ \hline
\small Data         &  \small $323$  \\ \hline
\end{tabular}
\captionof{table}{Expected and observed yields for $H\to WW$ selection.\label{tab:HWW_Base} }
\end{minipage}%

\end{figure}

\begin{figure}[ht]
\begin{minipage}[b]{0.5\columnwidth}%
    \centering
    \includegraphics[width=0.7\textwidth]{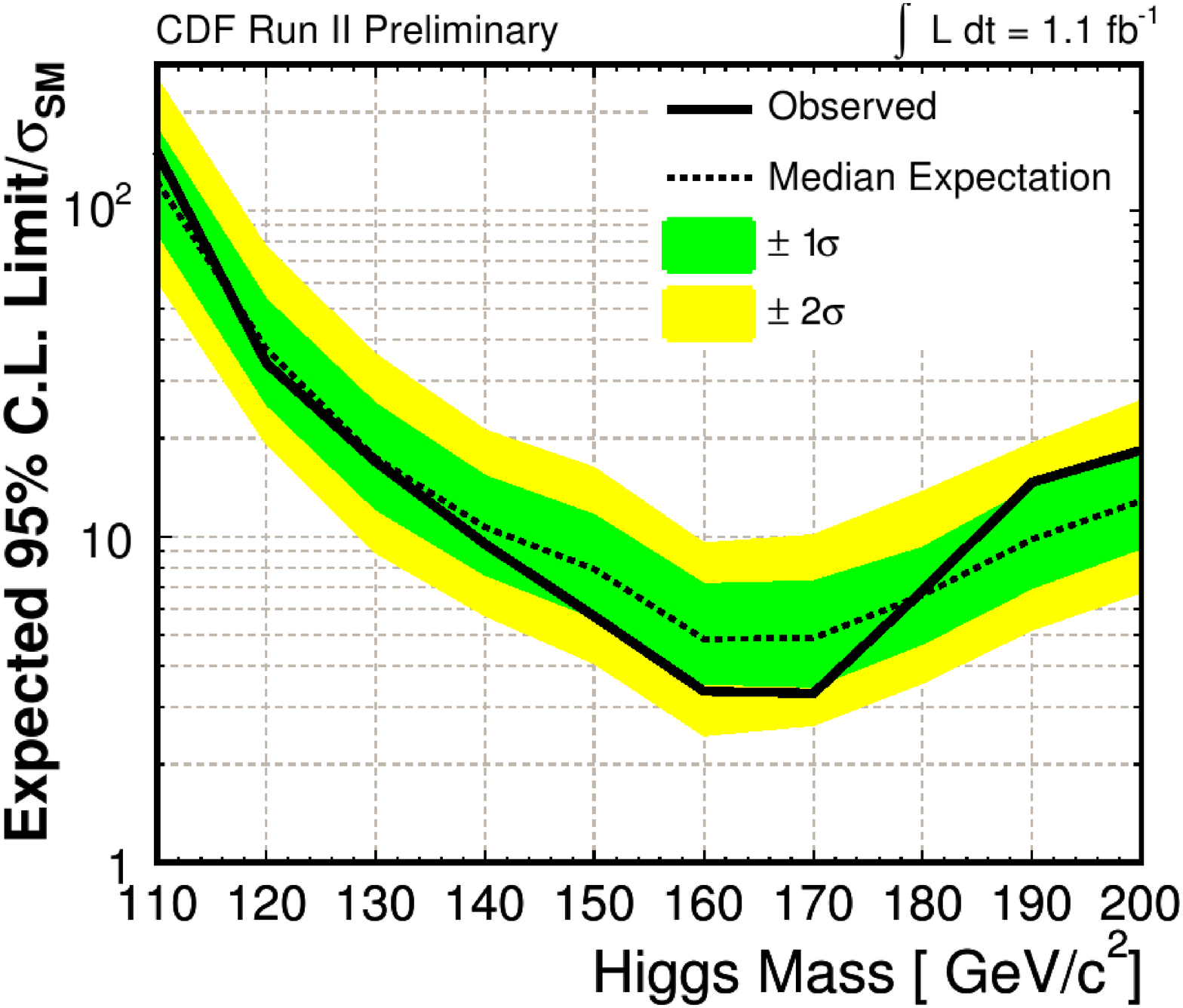}
    \caption{The ratio of 95\% C.L. upper limit of $H\rightarrow WW^{(*)}$ production to NNLL calculation 
    as a function of $m_H$.}\label{fig:HWW_Limit}
\end{minipage}%
\hfill%
\begin{minipage}[b]{0.45\columnwidth}%
\small
\centering
\begin{tabular}{|c|c|c|c|}
\hline
        &  $N_{exp}$ &  $\sigma_{95\%}$($~\rm{pb}$) & $\sigma_{95\%}/\sigma_{SM}$ \\ \hline
 $ 110$ &   $0.2$  &  8.9(7.1) & 151.2(122.6) \\
 $ 120$ &   $0.6$  &  4.7(4.9) &  33.9(37.4) \\ 
 $ 130$ &   $1.4$  &  4.0(3.8) &  17.0(17.4) \\
 $ 140$ &   $2.4$  &  3.0(3.4) &   9.5(10.7) \\ 
 $ 150$ &   $3.2$  &  2.1(2.9) &   5.7(8.0) \\
 $ 160$ &   $3.9$  &  1.3(1.8) &   3.4(4.8) \\ 
 $ 170$ &   $3.9$  &  1.2(1.7) &   3.3(4.9) \\
 $ 180$ &   $3.3$  &  1.9(1.8) &   6.8(6.6) \\ 
 $ 190$ &   $2.4$  &  2.8(1.9) &   14.6(9.8) \\
 $ 200$ &   $2.0$  &  2.8(2.0) &   18.4(12.9) \\ \hline
\end{tabular}
\captionof{table}{The expected yields, $N_{exp}$, and the observed (median expected) 95\% C.L. upper limit
$\sigma_{95\%}$, for the $H \rightarrow WW^{(*)}$ search.\label{tab:HWW_Limit} }
\end{minipage}
\end{figure}

The expected yield from each of the contributing backgrounds and the observed total are shown in
Table \ref{tab:HWW_Base}  while the expected yield due to an SM Higgs is shown as a function of mass in Table \ref{tab:HWW_Limit}.	
In order to maximize the sensitivity, the sample is divided into two parts based on the
expected signal to background (S/B) ratio for lepton identification categories. The corresponding LR distributions
are shown in Figure \ref{fig:HWW_LR}.
The limit of Higgs production cross section  is 
evaluated by performing 
a Bayesian binned maximum likelihood fit.
All of the background normalizations are free parameters in the fit but constrained to their 
expectations with a set of Gaussian constraints considering all of the 
assumed correlations between the systematics uncertainties. The limits of Higgs production cross 
section times $WW^{(*)}$ decay branching ratio and their ratios to NNLL calculations ($\sigma_{SM}$)
are shown in Table \ref{tab:HWW_Limit} and Figure \ref{fig:HWW_Limit}.

%%%%%%%%%%%%%%%%%%%%%%%%%%%%%%%%%%%%%%%%%%%%%%%%%%%%%%%
%                     ZZ                               %
%%%%%%%%%%%%%%%%%%%%%%%%%%%%%%%%%%%%%%%%%%%%%%%%%%%%%%%
\section{$ZZ$ Results}

\begin{figure}[ht]

\begin{minipage}[b]{0.7\columnwidth}%
    \centering
    \includegraphics[width=0.475\textwidth]{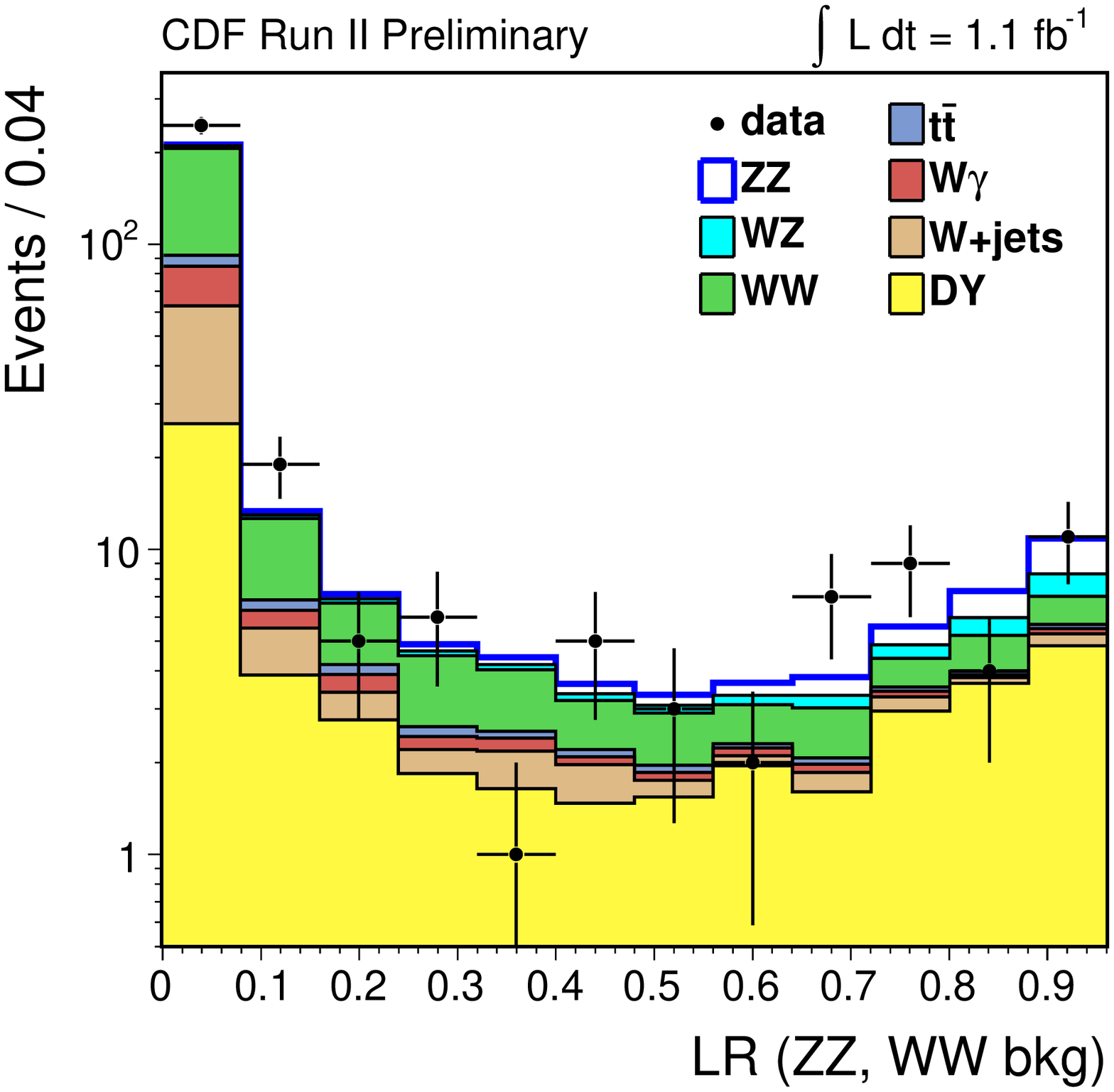}
    \unitlength=0.5\linewidth
    \put(-0.72,0.74){(a)}
    \includegraphics[width=0.475\textwidth]{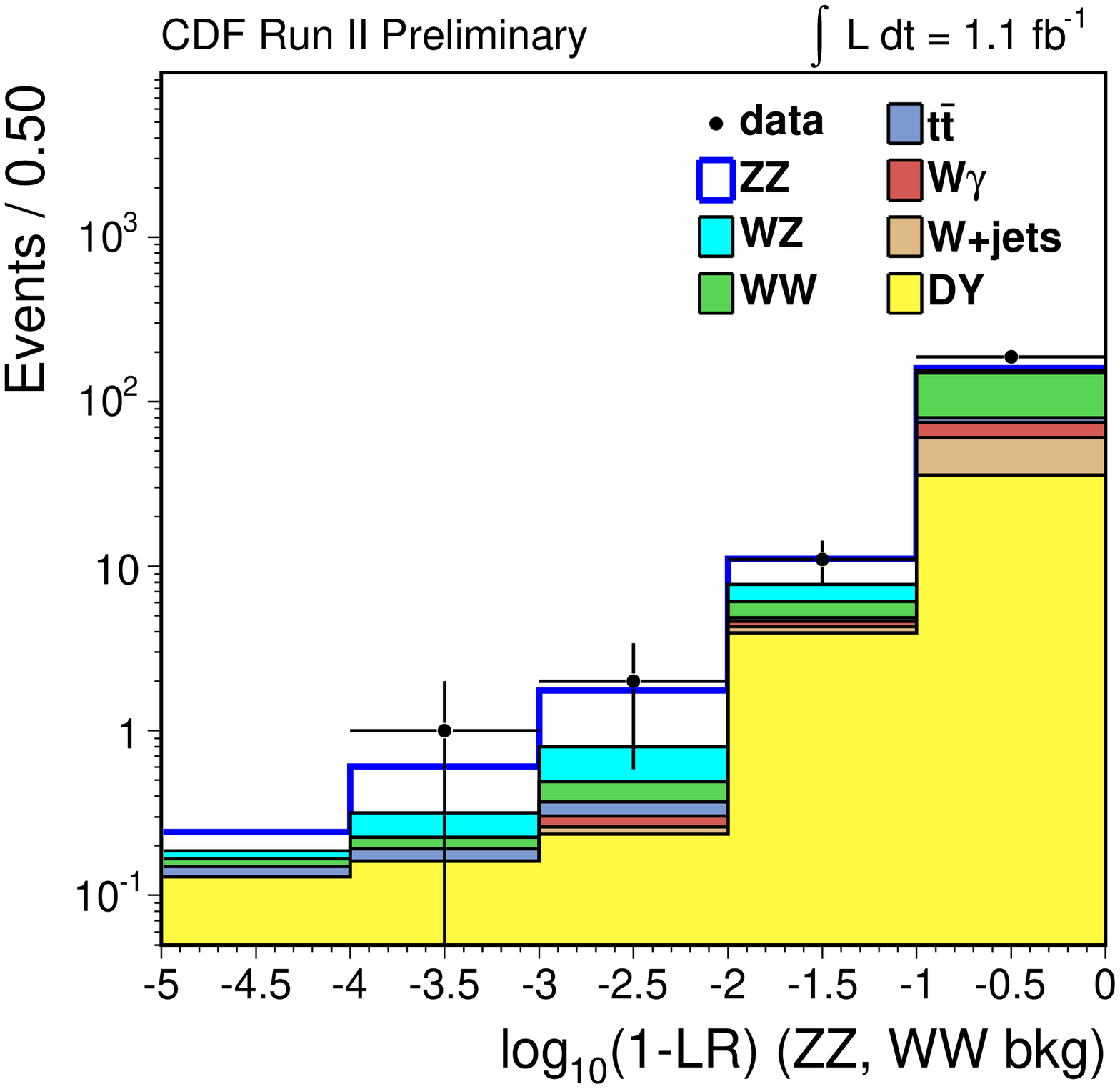}
    \put(-0.72,0.74){(b)}
    \caption{Distributions of (a) $LR={{P_{ZZ}}\over{P_{ZZ}+P_{WW}}}$ (b) $~\rm{log}_{10}(1-LR)$}\label{fig:ZZ_LR}
\end{minipage}%
\hfill%
\begin{minipage}[b]{0.25\columnwidth}%
\small
\begin{tabular}{|c|c|}
\hline
 &  Expected \\ \hline
 $ WW$        &   $ 69.2$ \\
 $ WZ$        &   $  7.1$ \\
 $ ZZ$        &   $ 10.7$ \\
 $ t\bar{t}$  &   $  5.1$ \\
 $ DY$        &   $ 24.0$ \\ 
 $ W\gamma$   &   $ 13.6$ \\
 $ W$+jets    &   $ 23.2$ \\ \hline
 Total        &   $152.9 \pm  11.6$ \\ \hline
 Data	      &   $  182$\\ \hline
\end{tabular}
\captionof{table}{Expected and observed yields for $ZZ$ selection.\label{tab:ZZ_Base} }
\end{minipage}%

\end{figure}

The expected and observed yields for the $ZZ$ selection are shown in Table \ref{tab:ZZ_Base}.
To avoid binning away information, the variable $log_{10}(1-LR)$
(shown in Figure \ref{fig:ZZ_LR}) is  used to set an upper limit.
The frequentist approach is used by performing background-only Monte Carlo experiments
based on the expected yields varied within the assigned systematics. For each experiment a test
statistic is formed from the difference in the log likelihood value with the background-only model 
and with the signal yield at the best fit value. The observed significance is 1.9$\sigma$ and we
set the 95\% CL upper limit of 3.4$~\rm{pb}$, which is consistent with the SM NLO cross section of $1.4\pm 0.1~\rm{pb}$.
This result has been combined with a search in the four charged lepton final state
to yield a total significance of 3.0 $\sigma$ \cite{Frank}.

%\begin{table}[hbt]
%\caption{Expected and Observed Yields for background in the Fit Region}
%\input{table/ZZ_yields_llnunu.tex}
%\begin{center}

%\begin{tabular}{|l|rrrrrrr|r|r|}
%\hline
%Category       &   $WW$        &   $WZ$        &   $ZZ$        &   $t\bar{t}$  &   DY          &   $W\gamma$   &   $W$+jets    &   Total       &   Data        \\
%\hline
%Total          &     69.2      &      7.1      &     10.7      &      5.1      &     24.0      &     13.6      &     23.2      &    152.9 $\pm$  11.6      &      182      \\
%\hline
%\end{tabular}
%\end{center}

%\label{tab:yield_ZZ}
%\end{table}

%%%%%%%%%%%%%%%%%%%%%%%%%%%%%%%%%%%%%%%%%%%%%%%%%%%%%%%
%                     Summary                         %
%%%%%%%%%%%%%%%%%%%%%%%%%%%%%%%%%%%%%%%%%%%%%%%%%%%%%%%
\section{Summary}

We have searched for a SM Higgs boson in the
$l^+l^-\MET$ final state with the Matrix Element method. 
The observed 95\% CL upper limit compares well with the expected upper limit as shown
in  Fig \ref{fig:HWW_Limit}. We see no sign of a significant excess or deficit
at any Higgs mass. The 95$\%$ CL upper limit for SM $ZZ$ production is  
3.4~\rm{pb} and consistent with the SM NLO cacluation.

\bibliography{moriond_schsu}% Produces the bibliography via BibTeX.

\begin{thebibliography}{1}

\bibitem{Catani:2003zt}
S.~Catani, D.~de~Florian, M.~Grazzini, and P.~Nason.
\newblock {\em JHEP}, 07:028, 2003.

\bibitem{LEPEWWG}
The lep electroweak working group, http://lepewwg.web.cern.ch/lepewwg/.

\bibitem{Campbell:1999ah}
J.~M. Campbell and R.~K. Ellis.
\newblock http://mcfm.fnal.gov/.
\newblock {\em Phys. Rev.}, D60:113006, 1999.

\bibitem{Frixione:2002ik}
Stefano Frixione and Bryan~R. Webber.
\newblock {\em JHEP}, 06:029, 2002.

\bibitem{Frank}
Frank W{\"u}erthwein.
\newblock Moriond qcd 2007, diboson production at the tevatron.

\end{thebibliography}

\end{document}